\documentclass[aps,pra,twocolumn,amsmath,amssymb,notitlepage]{revtex4-1}
\usepackage{graphicx}
\usepackage{cancel}
\usepackage{amsfonts}
\usepackage{amssymb}
\usepackage{amsthm}
\usepackage{url}

\usepackage{siunitx}
\usepackage{hyperref}
\usepackage{float}
\usepackage{multirow}  
\usepackage{dcolumn}   
\usepackage{bm}        
\usepackage{stmaryrd}  
\usepackage{enumerate}
\usepackage{stmaryrd}
\usepackage{ulem}

\usepackage{natbib}
\usepackage[dvipsnames]{xcolor}
\hypersetup{
	colorlinks,
	linkcolor={green!50!black},
	citecolor={blue!50!black},
	urlcolor={blue!80!black},
	citecolor   = {red!50!black} 
}
\def  \bsig    {\mbox{\boldmath$\sigma$}}

\begin{document}

\title{Determining the Rashba parameter from the bilinear magnetoresistance response in a two-dimensional electron gas}

\author{D.~C.~Vaz~$^{1,\star}$}
\altaffiliation[Current address:]{~CIC nanoGUNE BRT, Tolosa Hiribidea, 76, E-20018 Donostia – San Sebastian, Spain}
\author{F.~Trier~$^{1,\star}$}
\author{A.~Dyrda\l~$^{2,\star}$}
\email{adyrdal@amu.edu.pl}
\author{A.~Johansson~$^3$}
\author{K.~Garcia~$^{4,5}$}
\author{A.~Barth\'el\'emy~$^1$}
\author{I.~Mertig~$^3$}
\author{J.~Barna\'s~$^{2,6}$}
\author{A.~Fert~$^{1,4,5}$}
\author{M.~Bibes~$^{1}$}
\email{manuel.bibes@cnrs-thales.fr}

\address{$^1$~Unit\'e Mixte de Physique, CNRS, Thales, Universit\'e Paris-Saclay, 91767, Palaiseau, France\\
$^2$~Faculty of Physics, Adam Mickiewicz University in Pozna\'n, Uniwersytetu Pozna\'nskiego 2, 61-614 Pozna\'n, Poland\\
$^3$~Institut f\"ur Physik, Martin-Luther-Universität, Halle-Wittenberg, 06099 Halle (Saale), Germany\\
$^4$~CIC nanoGUNE BRT, Tolosa Hiribidea, 76, E-20018 Donostia - San Sebasti\'an, Spain\\
$^5$~Department of Materials Physics UPV/EHU, Apartado 1072, 20080 Donostia - San Sebasti\'an, Spain\\
$^6$~Institute of Molecular Physics, Polish Academy of Sciences, M. Smoluchowskiego 17, 60-179 Pozna\'n, Poland\\
$^\star$ These authors contributed equally to this work.}

\begin{abstract}
Two-dimensional (2D) Rashba systems have been intensively studied in the last decade due to their unconventional physics, tunability capabilities, and potential for spin-charge interconversion when compared to conventional heavy metals. With the advent of a new generation of spin-based logic and memory devices, the search for Rashba systems with more robust and larger conversion efficiencies is expanding. Conventionally, demanding techniques such as angle- and spin-resolved photoemission spectroscopy are required to determine the Rashba parameter $\alpha_{R}$ that characterizes these systems. Here, we introduce a simple method that allows a quantitative extraction of $\alpha_{R}$, through the analysis of the bilinear response of angle-dependent magnetotransport experiments. This method is based on the modulation of the Rashba-split bands under a rotating in-plane magnetic field. We show that our method is able to correctly yield the value of $\alpha_{R}$ for a wide range of Fermi energies in the 2D electron gas at the LaAlO$_{3}$/SrTiO$_{3}$ interface. By applying a gate voltage, we observe a maximum $\alpha_{R}$ in the region of the band structure where interband effects maximize the Rashba effect, consistently with theoretical predictions.
\end{abstract}

\date{\textcolor{RoyalBlue}{\today}}
\maketitle

\section{INTRODUCTION}

Magnetotransport phenomena in low-dimensional systems are extremely useful for applications in (nano)electronic devices. After a first generation of spintronic devices based on metallic magnetic multilayers \cite{Baibich1988,Binach1989,Camley89,Barnas90,Julliere,Dieny_in_Reig_Cardoso_book,Barnas_bookchapt2016}, the next generation exploits the idea of a full electrical control of the spin degree of freedom via electrically-induced spin torques of various origins. In this context, a key objective is to exploit spin-orbit coupling to achieve efficient charge-to-spin interconversion, owing to the spin Hall effect \cite{Dyakonov1971,Hirsch1999,engel07,sinova15} or the current-induced spin density also known as the inverse spin-galvanic effect, or alternatively as the Edelstein effect~\cite{Edelstein,aronov89,Ganichev2001,Ganichev2002}.

While in the beginning of the study of the Edelstein effect great emphasis was put on semiconducting heterostructures, owing to their strong bulk and structural inversion asymmetries \cite{Awschalom2007,Winkler_book,Fabian2007}, nowadays two dimensional (2D) systems show the most promise, with encouraging results found at the surfaces of topological insulators (TIs), two-dimensional graphene-like heterostructures and oxide interfaces \cite{Soumyanarayanan2016,Varignon2018,Song2018,Cummings2017,Han2014}. In these systems, the Rashba spin-orbit coupling acts as an internal momentum-dependent magnetic field that ensures a fixed relative orientation between the electron spin and momentum, an effect known as spin-momentum locking.

\begin{figure*}[ht!]
\centering
	\includegraphics[width=1\textwidth]{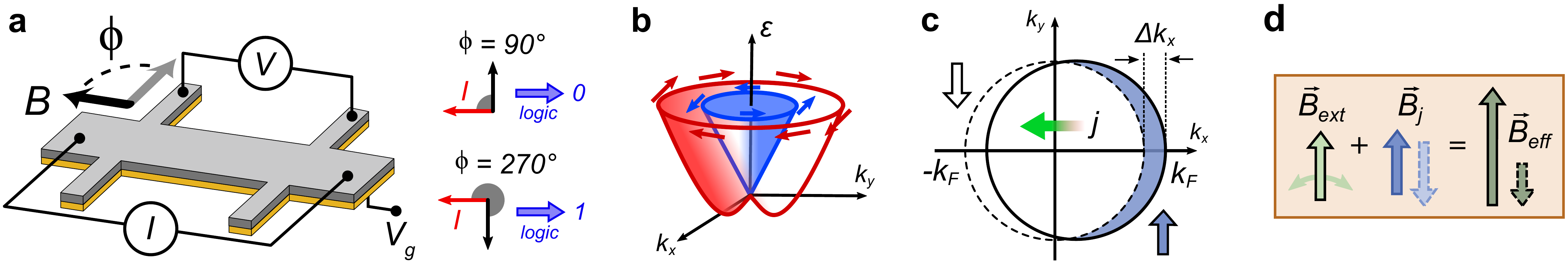}
	\caption{(a) Schematic of the BMR measurement setup under a rotating in-plane magnetic field $\mathbf{B}$. Two different resistive states are expected when $\mathbf{B}$ (black arrows) is perpendicular to the applied current $\mathbf{I}$ (red arrows). (b) Schematic of the electronic and spin structure of a 2D Rashba system. Arrows represent the electron spin. (c) Schematic of the shift of a Fermi contour by a bias current $j$, generating a non-equilibrium spin density $S$ (Edelstein effect \cite{Edelstein, aronov89, Ganichev2001, Ganichev2002}). The blue (white) arrow represents the orientation of accumulated (depleted) spins. (d) Effective magnetic field $\mathbf{B}_{eff}$ as a result of the current-induced magnetic field $\mathbf{B}_j$ (with its sign depending on the polarity of the applied current) and a rotating external field $\mathbf{B}_{ext}$ (adapted from \cite{guillet2020}).}
	\label{fig:BMR}
\end{figure*}

Recently, a new kind of magnetoresistance, the bilinear magnetoresistance (BMR), was reported in materials with strong spin-orbit coupling \cite{He2018,Dyrdal2019}. Here, two resistive states are observed depending on the relative orientation between the applied (bias) electric field and the magnetic field, as shown in Fig. \ref{fig:BMR}(a). In contrast with other recently reported unidirectional magnetoresistance effects (e.g., spin Hall magnetoresistance in multi-layered systems), the BMR appears in a single material. The effect was observed in topological insulators \cite{He2018} and in materials having surface or interface two-dimensional electron gases (2DEG) with Rashba interactions. For the latter, clear experimental evidence has been shown at the surface states of Ge(111), where the BMR in a magnetic field of 1T represented 0.5\% of the zero field resistance, and was larger than the magnetoresistance of standard symmetry \cite{guillet2020}. 

The BMR of the metallic surface states of Ge was interpreted by taking into account the locking $\alpha_R (\mathbf{k} \times \mathbf{z}) \cdot \bsig$ between spin $\bsig$ and momentum $\mathbf{k}$ in a Rashba 2DEG (Fig. \ref{fig:BMR}b). As shown in Fig. \ref{fig:BMR}c, the shift $\Delta k$ of the Fermi contour in the presence of a current $j$ and the resulting non-equilibrium energy $\Delta \varepsilon \sim \alpha_R (\Delta \mathbf{k} \times \mathbf{z}) \cdot \bsig$, with $\Delta k \sim j$, can be described by the introduction of a magnetic field $\mathbf{B}_j \sim \alpha_R (\mathbf{j} \times \mathbf{z})$ acting on the spin $\bsig$. The field $\mathbf{B}_j$ adds to the external field $\mathbf{B}_{ext}$, as shown in Fig. \ref{fig:BMR}d. Depending the direction of the current, $\mathbf{B}_j$ is added to (or subtracted from) the component of the applied field perpendicular to the current, and the BMR appears in cross-terms between $\mathbf{B}_{ext}$ and $\mathbf{B}_j$. Guillet et al. \cite{guillet2020} derived the Rashba parameter from the ratio between the BMR and the quadratic MR in $\mathbf{B}_{ext}$.

In this paper, we propose a microscopic theory of the BMR for Rashba 2DEGs. The mechanism is related to the existence of a current-induced magnetic field acting on the electron spins in systems with spin-momentum locking, and it dominates when the Rashba coupling is much larger than the Zeeman field.  We apply the model to the interpretation of experimental results in the LaAlO$_{3}$/SrTiO$_{3}$ (LAO/STO) system. As will be discussed, we develop the theory for the simple case of a single pair of circular Fermi contours characterized by a Rashba parameter. While in LAO/STO the BMR can be measured at different Fermi levels with different Fermi contours and Rashba splittings, the analysis shows a good agreement with results from tight-binding calculations even in the multiband regime. The fit of the model with the experimental BMR in a given voltage range thus provides us with an effective value of the Rashba parameter in the corresponding Fermi energy range. We demonstrate that the BMR signal may serve as a useful tool to characterize spin-to-charge interconversion efficiencies and to probe the strength of spin-orbit coupling in 2D systems. According to our proposition, an effective Rashba parameter can be determined solely based on the relative amplitudes of the quadratic and bilinear magnetoresistance terms, carrier density, and electron effective mass.

\section{THEORETICAL MODEL}

The electronic states of a two-dimensional Rashba gas are described by the following Hamiltonian in the plane-wave basis:
\begin{eqnarray}
 \label{eq:HR}
 \hat{H}_{R} = \frac{\hbar^{2} k^{2}}{2 m^{\ast}} \sigma_{0} + \alpha_{R} \left( k_{y} \sigma_{x} - k_{x} \sigma_{y} \right),
\end{eqnarray}
 where $m^{\ast}$ is the effective electron mass, $\alpha_{R}$ is the Rashba parameter, $k^{2} = k_{x}^{2}+k_{y}^{2}$, and $\sigma_{j}$ ($j= \left\{0, x, y, z\right\}$) are the unit Pauli matrices acting in spin space. The Hamiltonian (\ref{eq:HR}) above has two eigenvalues: ${\varepsilon_{\pm} = \frac{\hbar^{2} k^{2}}{2 m^{\ast}} \pm \alpha_R k}$.
 
\subsection{Semiclassical picture}
\label{sec:semiclass}

For an electric field in the $\hat{x}$-direction ($E_{x}$) and a relaxation time $\tau$, the momentum $\mathbf{k}$ acquires an extra component $\Delta k_{x} = e E_{x} \tau/\hbar$ ($e < 0$) which, from Eq. \ref{eq:HR}, leads to a non-equilibrium term $\alpha_{R} \Delta k_{x} \sigma_{y} = - \alpha_{R} e \tau E_{x} \sigma_{y}/\hbar$. This term is equivalent to an interaction with a current-induced Rashba field $\mathbf{B}_{j}$ along the $\hat{y}$-axis, as shown in Fig. \ref{fig:BMR}d, with $g \mu_{B} B_{j\,y} = b_{j\,y} = - \alpha_{R} e \tau E_{x}/\hbar$ (more generally, $\mathbf{b}_{j} = - \alpha_{R} e \tau (\hat{z} \times \mathbf{E})/\hbar$). Here, $\mathbf{b} = g \mu_{B} \mathbf{B}$ is given in energy units, where $\mathbf{B}$ is the magnetic field, $g$ the electron $g$-factor, and $\mu_{B}$ the Bohr magneton. It is also convenient to relate the field $b_{j\,y}$ to the non-equilibrium Edelstein spin polarization $S$ (see Fig. \ref{fig:BMR}c and Refs. \cite{Edelstein, aronov89, Ganichev2001, Ganichev2002}) induced by the current, $S_{y} = \frac{\alpha_{R} m^{\ast}}{2\pi \hbar} \Delta k_{x}$ for $\mathbf{E}$ along $\hat{x}$. A straightforward calculation leads to the relation $b_{j\,y} = \mathcal{J} S_{y}$, where $\mathcal{J} = - 2\pi \frac{\hbar}{m^{\ast}}$ and its sign changes for opposite signs of the effective mass $m^{\ast}$. (more details in Ref.~\cite{note2}).
 
\subsection{Microscopic description of BMR\\ in a Rashba 2DEG}
 
To calculate BMR we use the approach proposed recently by Dyrda\l~et al.~\cite{Dyrdal2019}. In the presence of external electric and magnetic fields, the total Hamiltonian can be written in the following form:
 \begin{eqnarray}
 \label{Htot}
 \hat{H}_{\mathbf{k}\mathbf{k}'}=  \left[\hat{H}^{0}_{\mathbf{k}}+ \hat{H}^{\mathbf{A}}_{\mathbf{k}}\right] \delta_{\mathbf{k} \mathbf{k}'} +\hat{V}^{\small{\rm imp}}_{\mathbf{k} \mathbf{k}'} \hspace{0.5cm}\\
 \hat{H}^{0}_{\mathbf{k}} =  \hat{H}_{R} + \hat{H}_{\mathbf{b}_{\rm eff}}\hspace{1.5cm}\\
\hat{H}_{\mathbf{b}_{\rm eff}} =  \mathbf{b} \cdot \bsig + \mathbf{b}_{j} \cdot \bsig \equiv \mathbf{b}_{\rm eff} \cdot \bsig\hspace{0.4cm}
 \end{eqnarray}
 
where $\hat{H}_{\mathbf{k}}^{\mathbf{A}} = - e\hat{\mathbf{v}} \cdot \mathbf{A}$ describes the coupling of the charge carriers to an external in-plane dynamical electric field. Here, $\hat{\mathbf{v}}$ is the velocity operator and $\mathbf{A}$ denotes electromagnetic vector potential, $\mathbf{A}(t) = \mathbf{A}_{\omega} \exp^{i \omega t/\hbar}$ with $\mathbf{A}_{\omega} = \frac{\hbar}{i \omega} \mathbf{E}_{\omega}$. The term $\mathbf{b} \cdot \bsig$ describes the coupling of the electron spin to an external in-plane magnetic field $\mathbf{b} = (b_{x}, b_{y})$. The term $\mathbf{b}_{j} \cdot \bsig$ expresses the coupling of the spin with the current-induced field described in Section \ref{sec:semiclass}. Without losing generality, one can assume that the external electric field is applied in the $\hat{x}$-direction. Thus, $\hat{H}_{\mathbf{k}}^{\mathbf{A}} = - e \hat{v}_{x} A_{x}$, and $\mathbf{b}_{j} \cdot \bsig = \mathcal{J} S_{y} \sigma_{y}$.

The relaxation occurs due to scattering from randomly-distributed point-like impurities. The potential $V(\mathbf{r})$ is assumed to be short-range with zero average ${\langle V(\mathbf{r})\rangle = 0}$, and the second statistical  moment ${\langle V(\mathbf{r}) V(\mathbf{r}')\rangle = w \delta(\mathbf{r} - \mathbf{r}')}$. For impurities distributed randomly at points $\mathbf{r}_{i}$, the potential ${V(\mathbf{r}) = \sum_{i} v(\mathbf{r} - \mathbf{r}_{i}) = \sum_{i} v_{0} \delta(\mathbf{r} - \mathbf{r}_{i})}$, where $v_{0}$ is the potential of a single impurity and $w = n_{i} v_{0}^{2}$. Thus, $\hat{V}^{\small{imp.}}_{\mathbf{k} \mathbf{k}'} = V_{\mathbf{k}\mathbf{k}'} \sigma_{0} $.

\subsection{Relaxation time and conductivity}

The relaxation time and conductivity have been calculated within the Green's function formalism.
The self-consistent Born approximation is used to find the relaxation rate $\Gamma$ (or relaxation time $\tau$). An important finding is that the relaxation rate/time is dependent on the external magnetic field and non-equilibrium spin density:
\begin{eqnarray}
\label{SE_final}
\Gamma(b) = \frac{\hbar}{2\tau_{b}} = \Gamma_{0} \left[  1 + 3  \left( \frac{ \mathcal{J} b_{y} S_{y}}{4 \Gamma_{0}^{2}} + \frac{b^{2}}{8 \Gamma_{0}^{2}}\right)\right] ,
\end{eqnarray}
where $\Gamma_{0} = n_{i} V_{0}^{2} \frac{m^{\ast}}{2 \hbar^{2}} \equiv \frac{\hbar}{2 \tau_0}$ is the relaxation rate in the absence of a magnetic field, $\Gamma_{0} = \Gamma(b = 0)$.

In turn, the longitudinal conductivity can be calculated from the expression~\cite{agd,mahan}:
\begin{equation}
\sigma_{xx} = \frac{e^{2} \hbar}{2 \pi}  \bigg \langle \mathrm{Tr} \int \frac{d^{2} \mathbf{k}}{(2\pi)^{2}} \hat{v}_{x} G_{\mathbf{k}}^{R} \hat{v}_{x} G_{\mathbf{k}}^{A} \bigg \rangle ,
\end{equation}
 where $\langle ... \rangle$ denotes the disorder average. In the simplest case, $\langle v_{x} G^{R} v_{x} G^{A} \rangle$ can be replaced by $v_{x} \bar{G}^{R} \mathcal{V}_{x} \bar{G}^{A}$, where $\bar{G}^{R,A}$ is the impurity averaged Green's function and $\mathcal{V}_{x}$ is the renormalized velocity related to the vertex correction.
Since the effective magnetic field is small and treated perturbatively, one can neglect its influence on the impurity vertex correction and consider the Hamiltonian of the 2DEG with only the Rashba term. In such a case, it is known that the impurity corrections to the velocity vertex function lead to cancellation of the so-called anomalous velocity, and $\mathcal{V}_{x} = \frac{\hbar k_{x}}{m^{\ast}} \sigma_{0}$ (see, e.g., Refs. \cite{Inoue,Dimitrova}).

Taking into account the impurity vertex correction and expanding Green's functions with respect to the effective magnetic field $\mathbf{b}_{\rm eff}$, we arrive at the final expression for the diagonal resistivity
\begin{equation}
\label{deltarho}
\rho_{xx} = \rho_{xx}^0 + \frac{3\pi}{4} \frac{h}{e^{2}} \left[ \frac{\alpha_R \tau_0}{|e| } \frac{j_{x} b \sin \phi}{\varepsilon_{R}^{2} + \varepsilon_F^{2}} + \frac{\varepsilon_F\tau_0}{h} \frac{b^{2} \cos(2\phi)}{\varepsilon_{R}^{2} + \varepsilon_F^{2}} \right],
\end{equation}
where $\rho_{xx}^{0} = \rho_{xx}(B = 0) = \frac{h}{e^{2}} \frac{\Gamma_{0} \varepsilon_{F}}{\varepsilon_{F}^{2} + \varepsilon_{R}^{2}}$, $\varepsilon_F$ is the Fermi energy, and $\varepsilon_{R} = \alpha_R k_{0}$ defines the Rashba energy (Rashba field) in the system (here $k_{0} = \sqrt{2 m^{\ast} \varepsilon_F}/\hbar$).
This expression clearly shows that even for a simple scalar scattering potential of point-like impurities, a bilinear magnetoresistance proportional to $j$ and $B$ appears, which reveals a $\sin(\phi)$ angular dependence with an oscillation period of 2$\pi$.

\subsection{Magnetoresistance}\label{magneto}

In the following we briefly discuss the behavior of the magnetoresistance $MR = \left[ \rho_{xx}(B) - \rho_{xx}(B=0)\right]$ obtained within the model under consideration. First, we extract the bilinear and $B$-quadratic (symmetric) components of the magnetoresistance following the definitions: ${BMR = \left[ MR(B, j_{x} = +j) - MR(B, j_{x} = -j)\right]/2}$ and ${QMR =  \left[ MR(B, j_{x} = +j) + MR(B, j_{x} = -j)\right]/2}$. Thus, the bilinear and quadratic magnetoresistance can be expressed as follows:
\begin{eqnarray}
\label{eq:BMR}
BMR = A_{BMR} \frac{j_{x}}{j} \sin\phi ,\\
\label{eq:QMR}
QMR = A_{QMR}  \cos(2\phi) ,
\end{eqnarray}
with the amplitudes
\begin{eqnarray}
\label{eq:BMRamp}
A_{BMR}=\frac{3\pi}{4} \frac{h}{e^{2}}  \frac{g\mu_B}{|e|} \frac{\alpha_R \tau }{\varepsilon_{R}^{2} + \varepsilon_F^{2}}jB  ,\\
 \label{eq:QMRamp}
A_{QMR}
= \frac{3\pi}{4} \frac{(g \mu_{B})^2}{e^2} \frac{\varepsilon_F\tau  }{\varepsilon_R^2+\varepsilon_F^2}B^{2}.
\end{eqnarray}
When we use the normalized magnetoresistance,  $MR = \left[ \rho_{xx}(B) - \rho_{xx}(B=0)\right]/\rho_{xx}(B=0)$,  then Eqs.~(\ref{eq:BMR}) and (\ref{eq:QMR}) still hold, but the amplitudes of BMR and QMR take the forms:
\begin{eqnarray}
\label{eq:BMRamp2}
A_{BMR}= \frac{3}{2} \pi \frac{g \mu_{B}}{|e| \hbar}\frac{\alpha_R \tau^{2}}{\varepsilon_{F}} j B ,\\
\label{eq:QMRamp2}
A_{QMR} = \frac{3}{4} \left(\frac{g \mu_{B}}{\hbar} \right)^{2} \tau^{2} B^{2}.
\end{eqnarray}

From Eqs.~(\ref{eq:BMR}) and (\ref{eq:QMR}) follows that the QMR oscillates with the periodicity of $\pi$ (as it is observed in usual anisotropic magnetoresistance experiments), whereas the BMR oscillates with the periodicity of $2\pi$. This behaviour results in a well pronounced asymmetry between $\phi =\pi /2$ and $\phi =3\pi /2$ in the angular dependence of the total magnetoresistance, as well as in an asymmetry of the total magnetoresistance for currents flowing in opposite directions. From Eq.~(\ref{eq:QMRamp2}), the amplitude of the QMR is expected to scale quadratically with an external magnetic field, a trend governed by the relaxation time $\tau$ in the system. From Eq.~(\ref{eq:BMRamp2}), the amplitude of the BMR is expected to scale linearly with both applied current and external magnetic field. Here, besides the dependence on external stimuli, $A_{BMR}$ also depends on material-dependent parameters, such as the relaxation time $\tau$, the Rashba parameter $\alpha_{R}$ and the Fermi energy $\varepsilon_{F}$. Lastly, the ratio between both amplitudes:
\begin{eqnarray}
\label{eq:ratio}
\frac{A_{BMR}}{A_{QMR}} \equiv \Lambda_{CS} = \frac{2 \pi \hbar}{|e| g \mu_{B}} \frac{\alpha_{R}}{\varepsilon_{F}} \frac{j}{B}
\end{eqnarray}
gives a $\tau$-independent relation from which $\alpha_{R}$ can be found experimentally, provided that $\varepsilon_{F}$ and $g$ are known. Here, this ratio is expressed by universal constants, the ratio of externally-controlled parameters, $j$ and $B$, and $\alpha_{R}$. Thus, this ratio, directly proportional to $\alpha_R$, is a quantity characterizing the Rashba coupling. Moreover, even in the case of a more complex multiorbital band structure strongly modified by spin-orbit interactions, the experimentally derived $\Lambda_{CS}$ can still provide a reasonable estimation of the magnitude of spin-orbit related effects, where an effective Rashba parameter may be considered [$\alpha_{R} \to \alpha_{eff}$ in Eqs.~(\ref{eq:BMRamp2}) and (\ref{eq:ratio})]. However, in this regime the model has to be used with caution, since a simple Rashba model cannot properly describe highly anisotropic Fermi contours. In the following section we describe why this approximation is reasonable for the system studied.

\section{EXPERIMENTAL RESULTS}

\begin{figure}[t]
\centering
	\includegraphics[width=1\columnwidth]{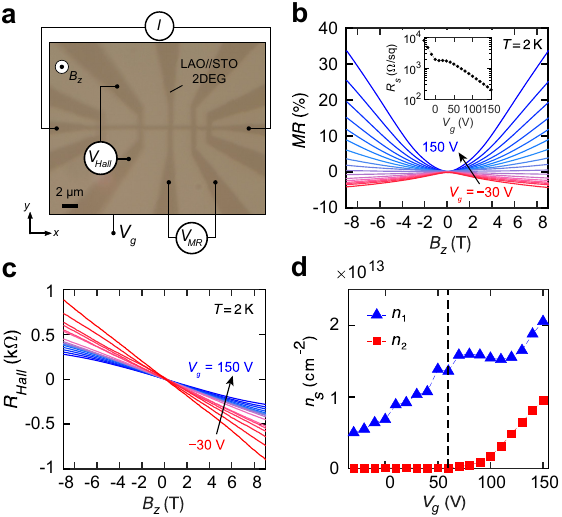}
	\caption{(a) Optical microscope image of a LAO/STO Hall bar device (b) Magnetoresistance ($MR$) and sheet resistance $R_{S}$ (inset) measurements as a function of the back-gate voltage $V_g$. (c) Hall measurements as a function of $V_g$. (d) Carrier densities $n_1$ and $n_2$, extracted from a two-band model fit of the Hall curves (considering different effective electron masses). }
	\label{fig:transport}
\end{figure}

We have performed angle-dependent transport experiments in the prototypical Rashba 2DEG found at the interface of LaAlO$_{3}$/SrTiO$_{3}$ (LAO/STO) oxide heterostructures. Hall bar devices [see Fig. \ref{fig:transport}(a)] were fabricated on LAO(1 unit cell)//STO through a combination of electron-beam lithography and a room-temperature deposition of a-LAO (30 nm) using pulsed laser deposition (more details in Ref. \cite{Trier2019}). As shown in Fig. \ref{fig:transport}(b), at low temperatures the device exhibits a positive magnetoresistance in the doped-regime (applying a back-gate voltage of $V_g$ = 150 V) and a negative magnetoresistance in the depleted-regime ($V_g$ = -30 V), signalling the well-known transition between weak antilocalization and weak localization regimes \cite{Caviglia2010}. Within this range of gate voltages, the sheet resistance $R_s$ increases from 200 $\Omega /\square$ to 8 k$\Omega /\square$ [inset of Fig. \ref{fig:transport}(b)]. Hall effect measurements as a function of $V_g$, shown in Fig. \ref{fig:transport}(c), confirm the n-type metallic behaviour of the device, similarly to unpatterned samples. By fitting these curves with a model assuming two types of charge carriers (with distinct effective mass) contributing to the conduction, we are able to extract the carrier densities $n_1$ and $n_2$, as shown in Fig. \ref{fig:transport}(d). For $V_g$ $\textless$ 60 V, linear Hall curves are obtained, indicating that only one band contributes to transport. In this range, the carrier density can be tuned between about 0.5 and 1.5 $\times$ 10$^{13}$ cm$^{-2}$. Above this $V_g$, heavy electron subbands start to be populated, resulting in slightly non-linear Hall curves and a maximum total carrier density of 3 $\times$ 10$^{13}$ cm$^{-2}$ (at $V_g$ = 150 V).

\begin{figure*}[ht!]
\centering
	\includegraphics[width=1\textwidth]{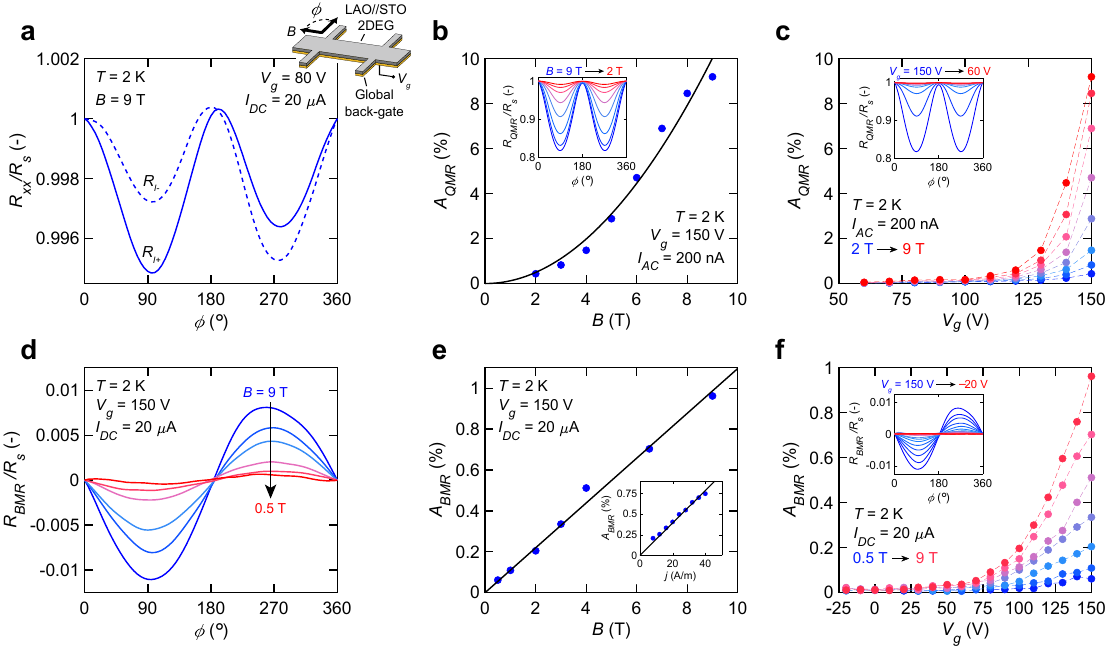}
	\caption{(a) Normalized longitudinal resistance under an in-plane rotating magnetic field $B$ for positive (continuous) and negative (dashed) DC bias current. (b) Amplitude of the first harmonic contribution (angle-dependence in the inset) as a function of $B$. Here, an AC current was used since QMR is independent of the current direction. This quadratic/symmetric part is given by $A_{QMR} = \{[R_{QMR}(\phi = \ang{0})-R_{QMR}(\phi = \ang{90})]/R_s\} \times 100\%$ and $R_{QMR} = (R_{I+}+R_{I-})/2$. (c) $A_{QMR}$ as a function of a back-gate voltage for different $B$ (angle-dependence in the inset). (d) Angle-dependence of the asymmetric contribution for different $B$, where $R_{BMR} = (R_{I+}-R_{I-})/2$. (e) Amplitude of the bilinear/asymmetric part as a function of $B$, given by $A_{BMR} = [(|R_{BMR}(\phi = \ang{90})|+|R_{BMR}(\phi = \ang{270})|)/2 R_s] \times 100\%$. Inset: $A_{BMR}$ as a function of the applied current density $j$. (f) $A_{BMR}$ as a function of a back-gate voltage for different $B$ (angle-dependence in the inset).}
	\label{fig:experiment}
\end{figure*}

We now move to the angle-dependent magnetoresistance experiments. We start by measuring the longitudinal resistance under a rotating in-plane external magnetic field $B$, as shown in Fig. \ref{fig:experiment}(a). At $\phi$ = $\ang{0}$, $B$ lies parallel to the applied DC current $I_{DC}$. Under $B$ = 9 T and $V_g$ = 80 V, the normalized magnetoresistance oscillates with a periodicity of 2$\pi$, in agreement with Eq.~(\ref{eq:QMR}). For opposite polarities of $I_{DC}$, a pronounced asymmetry between $\phi$ = \ang{90} and $\phi$ = \ang{270} is observed, a signature of the unidirectional and bilinear response predicted by our model and previously reported in the Rashba 2DEGs at LAO/STO interfaces \cite{Narayanapillai2014,Choe2019}, InAs quantum wells \cite{Choi2017} and Ge(111) \cite{guillet2020}. We performed similar angle-dependent measurements at different magnetic fields and gate voltages values and extracted the BMR and QMR as described earlier. We plot in Fig. \ref{fig:experiment}(b) $A_{QMR}$ as a function of the external magnetic field at $V_g$ = 150 V (angle-dependence in the inset). The data can be fitted with Eq.~(\ref{eq:QMRamp2}) (solid black line), confirming the quadratic dependence of $A_{QMR}$ with $B$. From the fitting, we extract a relaxation time of $\tau \sim$ 1 ps, consistent with previous reports in this system within the same carrier density range \cite{Caviglia2010}. Next, we evaluate $A_{QMR}$ as a function of $V_g$ at different external magnetic fields [Fig. \ref{fig:experiment}(c)]. By gating our device, the mobilities increase non-monotonically by one order of magnitude, leading to a similar modulation of the relaxation time, given that $\tau =(m^{\ast} \mu)/|e|$ (for carriers with a fixed effective mass $m^{\ast}$). From Eq.~(\ref{eq:QMRamp2}), $A_{QMR}$ is expected to scale with $\tau^2$, which explains its abrupt increase for larger mobilities at higher gate voltages.

In Fig. \ref{fig:experiment}(d), we show the asymmetric (or bilinear) part as a function of $B$. As shown in Fig. \ref{fig:experiment}(e), the amplitude of this asymmetric part $A_{BMR}$ is linear with both $B$ and the applied current density $j$, thus confirming its bilinearity. $A_{BMR}$ is also observed to strongly increase with applied $V_g$, as shown in Fig. \ref{fig:experiment}(f). However, unlike the gate dependence of $A_{QMR}$ and according to Eq.~(\ref{eq:BMRamp2}), $A_{BMR}$ depends not only on $\tau$ but also on the Fermi energy $\varepsilon_{F}$ and the Rashba parameter $\alpha_R$, rendering its analysis less trivial.

\begin{figure}[h]
\centering
	\includegraphics[width=1\columnwidth]{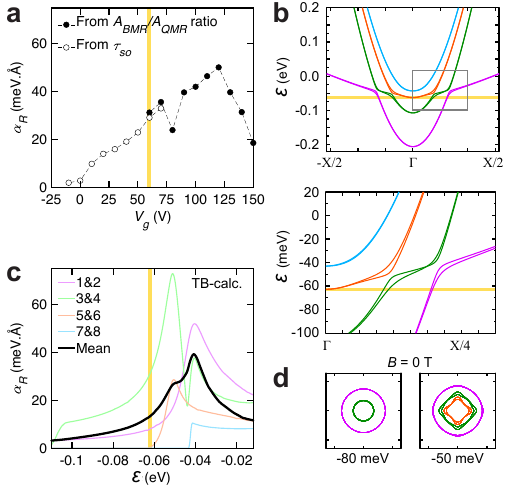}
	\caption{(a) Rashba parameter as a function of the back-gate voltage, extracted using the $A_{BMR}/A_{QMR}$ ratio from Eq.~(\ref{eq:ratio}) (full circles) and the $A_{BMR}$ from Eq.~(\ref{eq:BMRamp2}) with $\tau_{so}$ from Ref. \cite{Caviglia2010} (open circles). (b) Top: Band dispersion along the [100] direction, calculated through an eight-band tight-binding model. Bottom: Zoomed area with pronounced enhancement of the Rashba splitting. (c) Energy dependence of the Rashba parameter, calculated for each pair of Rashba-split bands (colors match the bands in panel b) and for its mean contribution weighted by the conductivity of each band pair (black line). Note that for the orange band pair $\alpha_R$ has a negative sign, not shown here for simplicity. (d) Fermi contours below ($\varepsilon_F$ = - 80 meV) and above ($\varepsilon_F$ = -50 meV) the Lifshitz transition. The yellow vertical lines in (a) and (c) and horizontal lines in (b) correspond to the Lifshitz transition.}
	\label{fig:theory}
\end{figure}

We use Eq.~(\ref{eq:ratio}) to extract $\alpha_R$ from the ratio $A_{BMR}/A_{QMR}$, $\varepsilon_F$ and the $g$-factor. The following results at different $V_g$ correspond to different Fermi energies, different bands, different Fermi contours and different Rashba parameters, as will be discussed further ahead. Thus, the interpretation of the BMR by our model in a given range of gate voltages characterizes an average value of the Rashba parameter for the corresponding Fermi contours. We start by calculating $A_{BMR}/A_{QMR}$. Since the $A_{QMR}$ fitting was only possible down to $V_g$ = 60 V, below this gate voltage $\alpha_R$ was derived using Eq.~(\ref{eq:BMRamp2}) with $\tau_{so}$ from magnetoresistance fittings to the Maekawa-Fukuyama formula in Ref. \cite{Caviglia2010}, which show a good agreement when compared with $\tau_{QMR}$ calculated using Eq.~(\ref{eq:QMRamp2}) up to $V_g$ $\approx$ 100 V. $\varepsilon_F$ can be estimated with $\varepsilon_F = (\hbar^2 \pi n)/ m^{\ast}$ for single parabolic bands. We note that, together with $g$, the estimation of $\varepsilon_F$ is the biggest source of error when implementing this model, since it requires previous knowledge of $m^{\ast}$. Nevertheless, using a fixed $m^{\ast}$ = 1.3 $m$ ($m$ is the electron mass), $g$ = 0.5 (from Ref. \cite{Caviglia2010}), and the carrier densities showed in Fig. \ref{fig:transport}(d), we plot in Fig. \ref{fig:theory}(a) the calculated $\alpha_R$ as a function of $V_g$. We observe that $\alpha_R$ increases from 2 meV\si{\angstrom} up to 50 meV\si{\angstrom} at $V_g$ = 120 V, where it reaches its maximum value. Beyond this, $\alpha_R$ rapidly decreases, reaching a value of 20 meV\si{\angstrom} at $V_g$ = 150 V. This nonmonotonic behaviour is reminiscent of recent results in SrTiO$_3$-based 2DEGs, where the spin-to-charge current conversion efficiency \cite{Vaz2019} and spin current generation and detection \cite{Trier2019} were modulated up and down using gate voltages. The values extracted are in good agreement with the $\alpha_R$ reported in other studies for the LAO/STO system in similar carrier density regimes \cite{Caviglia2010,King2014}. Here, we also highlight that, contrary to the other methods, our model is applicable to a broader range of carrier densities, especially where $\alpha_R$ takes the highest values.

We further clarify the origin of this modulation by deriving values of $\alpha_{R}$ from an eight-band model Hamiltonian (four Rashba-split bands with opposite spin orientation) composed of two light $d_{xy}$ bands (lower in energy) and one heavy $d_{yz}$ and $d_{xz}$ band (higher in energy), as depicted in Fig. \ref{fig:theory}(b). Details of the tight-binding modelling can be found in Ref. \cite{Vaz2019}. We start by calculating $\alpha_{R} = \Delta k(\hbar^2 / 2m^{\ast})$ directly from the energy spectrum, where $\Delta k= \overline{k}_{outer}-\overline{k}_{inner}$ gives the difference of two neighbouring subbands. In order to respect the individual contribution of each band pair to the QMR and BMR signals, the mean value $\overline{\alpha}_R$ is calculated as an average of all $\alpha_R$ from each band pair weighted according to their contribution to the charge conductivity. We observe in Fig. \ref{fig:theory}(c) that $\overline{\alpha}_R$ increases from 5 meV\si{\angstrom} to a maximum of 40 meV\si{\angstrom} at around $\varepsilon_F$ = -40 meV, coinciding with the crossing between the heavy $d_{xz,yz}$ and light $d_{xy}$ subbands \cite{Joshua2012,King2014}, followed by a sharp decrease similar to what was observed experimentally in Fig. \ref{fig:theory}(a). We can identify a critical energy $\varepsilon_\mathrm{c}$ (or critical $V_g$) up to which the splitting of the subbands may be directly estimated by a simple Rashba spin-orbit coupling, characterized by circular Fermi contours [left panel in Fig. \ref{fig:theory}(d)]. For our system, $\varepsilon_{\mathrm{c}}$ $\approx$ -60 meV, where heavy $d_{xz,yz}$ subbands start to be populated (also known as Lifshitz transition, represented with a yellow line in Fig. \ref{fig:theory}). As follows from the TB model results in Fig. \ref{fig:theory}(c), below this Lifshitz point $\alpha_{R}$ is mostly determined by the electronic states originating from the first pair of subbands. In this range, $\alpha_R$ is sufficient to properly describe the Rashba system, and our analytical formulas derived for a simple Rashba model fit well to the experimental data. Above the Lifshitz point the energy spectrum becomes more complex, as displayed in the right panel of Fig. \ref{fig:theory}(d), and $\alpha_{R}$ can no longer be interpreted as a simple Rashba parameter, but rather as an approximation of the spin-orbit coupling strength in the system, given by $\alpha_{eff}$ (as described in Section \ref{magneto}). Although in this regime the complex nature of the Fermi contours should in principle prevent us from using our model, we note that the majority of the conductivity above the Lifshitz point is in fact carried by the first pair of subbands, which has the highest density of states. Consequently, in Fig. \ref{fig:theory}(c)  we observe that the magenta line (bands 1\&2) has a very large weight on the mean contribution (black curve). We conclude that the calculated $\alpha_{eff}$ in this range can be approximated to the behaviour of simple Rashba bands, which gives us a fair comparison with respect to the experimental data.

\section{CONCLUSION}

In summary, we have developed a model of bilinear magnetoresistance in Rashba systems, based on the  existence of a current-induced magnetic field that acts on the electron spins in systems with spin-momentum locking. We employed this model to study the angle-dependent magnetoresistance measured in the Rashba 2DEG at the LAO/STO interface and derived the full gate dependence of $\alpha_{R}$. A maximum $\alpha_{R}$ is observed at $V_g \approx$ 120 V, coinciding with the crossing between light and heavy subbands, and – importantly – where other magnetotransport probes of Rashba physics are not applicable. Lastly, given the increasing interest of large spin-orbit coupling 2D systems for spin logic and computational applications \cite{intel2019,noel2020}, this model provides a useful tool to study new Rashba systems with potentially larger $\alpha_R$, which may exhibit giant spin-to-charge current conversion or spin-orbit torque effects.

{\textit{Acknowledgement.}} -- This work received support from the ERC Consolidator Grant 615759 “MINT”, ERC Advanced Grant 833976 “FRESCO” and partial support by the National Science Center in Poland as a research project No. UMO-2018/31/D/ST3/02351. A.~D. acknowledges support from the French Embassy through the French Government Scholarship (BGF 2019). A.~J. acknowledges  support from grant CRC/TRR 227 of Deutsche Forschungsgemeinschaft (DFG). We also thank J.-M. George for insightful discussions.


\begin{thebibliography}{99}
\bibitem{note2}
Though the current-induced Edelstein spin polarization $\mathbf{S}$ is related to the field $\mathbf{b}_j$, the latter can be calculated quasiclassically without taking into account the spin polarization. This is because both $\mathbf{b}_j$ and $\mathbf{S}$ are proportional to the momentum shift $\Delta \mathbf{k}$. 
Due to its mathematical form, $\mathbf{b}_{j}\cdot \bsig \sim \mathbf{S} \cdot \bsig$ is sometimes referred to as effective exchange or effective magnetic field. However, one should bear in mind that the origin of $\mathbf{b}_{j}$ is the spin-orbit interaction which has a distinct physical nature to the exchange interaction of $\bsig$ with $\mathbf{S}$.
\end{thebibliography}
\end{document}